\documentclass[aps, superscriptaddress,   twocolumn
]{revtex4-1}

\usepackage{latexsym,amsfonts,amssymb,exscale,enumerate,comment, bbold}
\usepackage{amsmath,amsthm,amscd}
\usepackage[all,knot]{xy}
\usepackage{tikz}
\usetikzlibrary{decorations.markings}
\usetikzlibrary{decorations.pathreplacing}
\tikzstyle directed=[postaction={decorate,decoration={markings,
    mark=at position #1 with {\arrow{>}}}}]

\usepackage{braket}

\usepackage{url}
\usepackage[bookmarks=true,%
    colorlinks=true,%
    linkcolor=blue,%
    citecolor=blue,%
    filecolor=blue,%
    menucolor=blue,%
    urlcolor=blue,%
    breaklinks=true]{hyperref}

\usepackage{graphicx}
\usepackage{color}

\theoremstyle{plain}

\theoremstyle{definition}

\theoremstyle{definition}

\newcommand{\hackcenter}[1]{
 \xy (0,0)*{#1}; \endxy}

\renewcommand{\to}{\rightarrow}

\def\Id{\mathrm{Id}}

\def\mf{\mathfrak}

\numberwithin{equation}{section}
\let\tilde=\widetilde

\let\epsilon=\varepsilon

\usepackage{bbm}
\def\C{{\mathbb{C}}}

\def\R{{\mathbb R}}
\def\Z{{\mathbb Z}}
\def\Q{{\mathbb Q}}

\def\1{\mathbbm{1}}%
\usepackage{bm}

\allowdisplaybreaks

\usepackage[normalem]{ulem}

\begin{document}
\title{Universal quantum computation using Ising anyons from a non-semisimple Topological Quantum Field Theory}

\author{Filippo Iulianelli}
 \affiliation{Department of Physics, University of Southern California, Los Angeles, California 90089,USA}
\author{Sung Kim}
\affiliation{Department of Mathematics,
 University of Southern California,
  Los Angeles, California 90089, USA}
\author{Joshua Sussan}
\affiliation{Department of Mathematics,
  CUNY Medgar Evers,
  Brooklyn, NY 11225, USA}
  \affiliation{Mathematics Program,
 The Graduate Center, CUNY,
  New York, NY 10016, USA}

\author{Aaron D. Lauda *}
\affiliation{Department of Mathematics,
 University of Southern California,
  Los Angeles, California 90089, USA}
  \affiliation{Department of Physics, University of Southern California, Los Angeles, California 90089,USA}

\begin{abstract}
We propose a framework for topological quantum computation using newly discovered non-semisimple analogs of topological quantum field theories in 2+1 dimensions. These enhanced theories offer more powerful models for quantum computation. The conventional theory of Ising anyons, which is believed to describe excitations in the $\nu = 5/2$ fractional quantum Hall state, is not universal for quantum computation via braiding of quasiparticles. However, we show that the non-semisimple theory introduces new anyon types that extend the Ising framework. By adding just one new anyon type, universal quantum computation can be achieved through braiding alone.  This result opens new avenues for realizing fault-tolerant quantum computing in topologically ordered systems.
\end{abstract}

\maketitle
\setcounter{tocdepth}{3}

%
 
\section{Introduction}
%
Topological quantum computation (TQC) has been proposed as a framework for quantum computation with the potential for scalability and fault-tolerance~\cite{Kit, FKLW}.    By encoding logical qubits in non-local topological characteristics of a system, TQC protects against decoherence arising from local interactions.
 Such theories rely on excitations, or quasiparticles called anyons, that can exist in 2D many-body systems.  In many instances, such quasiparticles are believed to obey more exotic exchange statistics extending properties of the usual bosons and fermions.   These anyons are of particular interest for topological quantum computation because they offer a robust platform for encoding and manipulating quantum information.

Transporting one anyon adiabatically around another induces a unitary transformation on the ground state Hilbert space that depends only on the path used to carry out the exchange.  If this transformation is more than a simple phase, we say that the anyons are non-abelian.  Systems supporting non-abelian anyons can give rise to models for quantum computation where qubits are encoded into configurations of non-abelian anyons and quantum gates are performed by braiding anyons around each other.   Large classes of examples are known to provide universal quantum computation~\cite{FKLW,FLW02}.

Fractional quantum Hall (FQH) systems are a physical framework that offer a candidate platform for topological quantum computation.
The fractional quantum Hall state in the second Landau level with filling fraction $\nu=5/2$ is the most promising two-dimensional system believed to support non-Abelian anyons.     
This filling fraction $\nu=5/2$ has been experimentally observed~\cite{PhysRevLett.59.1776}, and Moore and Read~\cite{MOORE1991362} proposed a ground state wavefunction 
utilizing correlators coming from conformal field theory describing the Ising model. Their work  predicted that the elementary excitations of the $\nu=5/2$ state would be non-abelian~\cite{NAYAK1996529,PhysRevB.83.075303,PhysRevLett.90.016802}. This Moore-Read, or Pfaffian state \cite{GREITER1992567}, has been found numerically to approximate the $\nu=5/2$ FQH state remarkably well across a wide range of microscopic interactions~\cite{PhysRevLett.80.1505,PhysRevLett.84.4685}.

The non-abelian anyons arising from the Moore-Read state are called Ising anyons since the fusion of quasiparticles obeys the chiral Ising model fusion rules~\cite{MOORE1991362}. 
\begin{equation} \label{eq:Ising-fusion}
\begin{split}
\psi \times \1 &= \psi,   \quad 
\sigma \times \1 = \sigma, \quad
\psi \times \psi = \1, \\
\psi \times \sigma &= \sigma, \quad
\sigma \times \sigma = \1 + \psi.
\end{split}
\end{equation}
The anyon types consist of a fermion $\psi$, a $\sigma$ anyon, as well as the vacuum state $\1$. The last rule implies that the fusion of two $\sigma$-anyons has two possible outcomes consisting of either the vacuum or the $\psi$ anyon.  They are promising candidates for TQC due to their large quasiparticle gap that suppresses the creation of thermal particles that can interact with the system.  The nonzero electric charge of the anyons allows for electrostatic gates that are estimated to produce very low error rates~\cite{PhysRevLett.94.166802}.

While Ising anyons are one of the most promising candidates to be realized experimentally, the primary obstruction to utilizing Ising anyons as a framework for TQC is that the braiding of anyons can only produce Clifford gates, and hence are not universal for quantum computation~\cite{BRAVYI2002210,MR1910833}.  Some authors have suggested extended protocols beyond braiding alone.
In \cite{PhysRevB.73.245307} 
the authors show that including topologically unprotected operations enable the Pfaffian state at $\nu=5/2$ to support universal quantum computation, with some operations requiring error correction.  Similar results were obtained in \cite{PhysRevA.73.042313} utilizing short-range interactions of anyons.

Here we propose a new framework for Ising anyons that allows for universal TQC by braiding alone.  This new approach arises from the well-known interplay between TQC and topological quantum field theory (TQFT)~\cite{FKLW,GTKLT,Kit,LW}.  The classical paradigm connects the mathematical structure of a unitary modular tensor category used to construct TQFT with anyon models.   Recently, new 2+1-dimensional TQFTs have been discovered that produce much more powerful topological invariants~\cite{CGP1,BCGP1,CGP2}.  These are achieved by moving beyond the unitary modular tensor category (UMTC) framework by leveraging \emph{non-semisimple} categories.

In previous work, the last two authors and collaborators have proposed that the heightened topological power of the non-semisimple theories may lead to more powerful models of TQC~\cite{GLPMS2,GLPMS4}.  Here we show that this is the case.  From the TQFT perspective, Ising anyons arise from the representation theory of the quantum group associated to $\mf{sl}(2,\C)$ with the quantum parameter $q=e^\frac{\pi i}{4}$.  The construction of a UMTC from this data is somewhat subtle as it arises by first considering the category of representations and then performing a semisimplification process ~\cite{MR1091619,MR2286123} that throws out an infinite number of representations with so-called quantum trace zero.

The power of non-semisimple TQFT is that it bypasses the semisimplification process and harnesses the representations that would have been removed by renormalizing the quantum trace~\cite{GKP1}.   The result is no longer a modular tensor category in the traditional sense but still gives rise to robust topological invariants.

The non-semisimple TQFT framework also supports anyon theories. In the case of $\mf{sl}(2,\C)$ at an eight root of unity, we obtain anyons obeying similar fusion rules to Ising anyons, except now there is a new anyon type $\mathbf{\alpha}$ indexed by a real number. The symbol $\alpha$ denotes both the anyon type and the real number it is indexed by. Here, we show that by including just a single new anyon $\alpha$, universal quantum computation can be achieved by braiding alone where the $\alpha$-anyon remains fixed throughout the computation.

While the non-semisimple theories have a number of topological advantages, they possess one significant obstacle, namely, unitarity.  
In previous work, the last two authors along with Geer and Patureau-Mirand showed that the non-semisimple TQFTs gave rise to Hermitian theories that are unitary, but with respect to a possibly indefinite form~\cite{GLPMS}.  
Despite the indefinite unitarity in general, in this work, we give an encoding for a multiqubit model where the computational Hilbert space is positive definite and braiding alone can produce universal quantum computation.  Indefinite unitarity only appears in the noncomputational space. Our single qubit Hilbert space is positive definite for certain intervals of $\alpha$, and we show that braiding alone is dense in $SU(2)$.

For multiple qubits, braiding will, in general, cause leakage into the noncomputational space.  To construct low-leakage entangling gates, we follow an approach similar to \cite{PhysRevB.75.165310,PhysRevLett.96.070503,PhysRevLett.95.140503} developed in the context of Fibonacci anyons.  Since two Ising type anyons fuse into two channels $\sigma \times \sigma = \1 + \psi$, one of which is the vacuum, braiding a pair of $\sigma$ anyons through a collection of other anyons will have no effect if they fuse to the vacuum state and will perform a nontrivial unitary transformation when in the $\psi$ state.  In this way, we can perform controlled braiding operations to entangle two qubits.
Following Reichardt~\cite{Reichdist}, see also \cite{Cui-leak,GrifCui}, we give an explicit and efficient iterative procedure for constructing control phase gates with arbitrarily small leakage into the noncomputational space.  These techniques do not require the Solovay-Kitaev method or brute force beyond finding initializing braids with the desired properties.

Finally, despite the global indefinite unitarity in our setting, this theory may still be physically realizable.   There has been a resurgence of interest in non-Hermitian physical systems, especially within the context of topological and symmetry-protected phases.   Indeed, just as the traditional TQFTs have a close connection with rational conformal field theories~\cite{MR990772}, these new non-semisimple TQFTs are known to connect with logarithmic CFTs~\cite{MR2283660,MR3795642,Negron}. In the case we consider here, the relevant logarithmic vertex operator algebra is the singlet~\cite{MR2016652} vertex algebra $\mathcal{M}(p)$ for $p \in \Z_{\geq 2}$, which corresponds to $p=4$ central charge $c=13-6p- \frac{6}{p}$ in the relationship between quantum groups and logarithmic CFT. 

Logarithmic CFTs often have a similar lack of unitarity, despite providing effective descriptions of a variety of physical phenomena such as disordered systems~\cite{MAASSARANI1997603}, non-equilibrium processes, and boundary phenomena in statistical mechanics~\cite{SALEUR1992486,ROZANSKY1992461}, 
polymers and percolation~\cite{PhysRevLett.58.2325,Flohr,MR4785749}, and
two-dimensional turbulence~\cite{Tabar_1997}. 

 Furthermore, while some of the known logarithmic CFTs containing Ising anyons are gapless (such as the Haldane-Rezayi fractional quantum Hall state at $\nu=5/2$) we see evidence that the non-semisimple theories we develop are connected to gapped theories.  To see this, it is helpful to consider the strong parallels between the physical realizability of our model and that of Fibonacci anyons associated with $\nu = 12/5$ FQH states~\cite{PhysRevB.59.8084,Slingerland_2001}.

The standard Fibonacci anyon theory is a chiral topological phase that cannot be realized by a commuting projector Hamiltonian in strictly two dimensions~\cite{Kitaev_2012}. Instead, it is typically described by an effective Hamiltonian acting on the constrained fusion Hilbert space of multiple Fibonacci anyons, where dynamics are governed by braiding and fusion rules~\cite{NKK}. In contrast, the doubled Fibonacci theory, a closely related non-chiral topological phase, can be realized microscopically as the ground state of a Levin-Wen model~\cite{KKR,Simon:2023hdq}: a gapped, commuting projector Hamiltonian defined on a 2D lattice with local degrees of freedom and topological order~\cite{LW}.

These doubled models exhibit ground state degeneracy on higher-genus surfaces~\cite{Hu_2012} and support anyonic excitations with fusion and braiding governed by the Drinfeld center of the input category~\cite{LW,KKR}. Mathematically, the chiral and non-chiral theories correspond to the Witten–Reshetikhin–Turaev~\cite{MR1091619} and Turaev–Viro~\cite{TURAEV1992865} topological quantum field theories, respectively, with a well-understood relationship connecting them via the Drinfeld center construction~\cite{Tu,KKR}.

In prior work, we introduced a broad class of non-semisimple Levin-Wen models ~\cite{GLPMS2,GLPMS3} that likewise admit microscopic descriptions via gapped, local commuting projector Hamiltonians.  These Hamiltonians are \textit{pseudo-Hermitian}, reflecting the non-semisimplicity of the input data. Even with the indefinite norms, pseudo-Hermitian theories give rise to a real spectrum of energy eigenvalues, normalizable wave functions, and time evolution by an exponential of the Hamiltonian that is self-adjoint in the indefinite norm~\cite{Most-rep}. 

These new non-semisimple Levin-Wen models support topological ground state degeneracy and anyonic quasiparticle excitations.   However,  as with the chiral (non-doubled) Fibonacci theory, the chiral phase we propose in this article does not currently have a known Hamiltonian realization. It may instead serve as an effective theory that captures the low-energy emergent degrees of freedom of a more complex microscopic model, in analogy with the effective Hamiltonian description of Fibonacci anyons~\cite{NKK}.


 The Hilbert space of our theory is also closely connected with the modified boundary conditions of the open XXZ spin chain studied in \cite{Chernyak_2022}.
%
 The connection with these condensed matter models suggests that FQH states supporting the new anyon types introduced here may be realizable within existing paradigms for realizing TQC experimentally, possibly by introducing novel boundary conditions. 

\section{Results} 

\subsection{Non-semisimple Anyons}

The fusion rules governing anyons in our theory are closely related to $SU(2)_k$ Chern-Simons theory at level $k=2$ in which each quasiparticle has a half-integer ``$q$-deformed"  spin quantum number, or $q$-spin, analogous to ordinary spin.
Our fusion rules arise from the unrolled quantum group for $\mf{sl}(2,\C)$ with the quantum parameter set to an eighth root of unity~\cite{GLPMS}. This theory contains a vacuum state $\1$, and anyon types $\psi$, $\sigma$ of $q$-spin 0, 1/2, and 1, respectively, as in the Ising theory,  but also possesses new `non-semisimple' anyon types removed in $SU(2)_2$ Chern-Simons theory.

The non-semisimple theory retains additional anyon types that would have been removed in the traditional semisimplification process for $SU(2)_2$.   These include new anyon types $\alpha$, indexed by non-half-integer real numbers, whose $q$-spins are no longer restricted to have half-integer values.  The Hilbert space for the theory we propose augments Ising anyons by a single $\alpha$-type anyon.  However, to fully understand the fusion rules and constraints on the associated $F$-symbols, we must also consider $q$-spin 3/2 and 2 particles $S_{3/2}$, and $P_2$ , all of which have traditional quantum trace zero at level $k=2$. 

The fusion rules relevant to our encoding are given by 
\begin{equation} \label{eq:alpha-fusion}
\begin{split}
\alpha \times \1 = \alpha,   \quad 
\alpha \times \sigma &= (\alpha + 1) + (\alpha - 1),\\
\sigma \times \sigma = \1 + \psi, \quad 
\alpha \times \psi &= (\alpha+2)+\alpha+(\alpha-2),\\
\sigma\times \psi =\sigma+S_{3/2}, \quad \psi \times \psi& =\1+P_2.\\ 
\end{split}
\end{equation}
Observe that removing the anyon types $S_{3/2}$ and $P_2$ in the last line, as in the semisimplification procedure, results in the more familiar Ising fusion rules \eqref{eq:Ising-fusion}.  A complete list of fusion rules can be found in \cite{CGP2}.

Consider the vector space $\mathcal{H}_{n} := \mathcal{H}_{\alpha; \sigma^{2n}}$ consisting  of a single type $\alpha$-anyon and $2n$ $\sigma$-type anyons in a total $q$-spin state $\alpha$.  The fusion rules \eqref{eq:alpha-fusion} imply that the space $\mathcal{H}_n$ is $\binom{2n}{n}$-dimensional and will contain our $n$-qubit topologically protected Hilbert space. The single qubit space $\mathcal{H}_1$ is encoded  as
$\ket{0} = ((\alpha,\sigma)_{\alpha+1}, \sigma)_{\alpha}$,
$\ket{1} = ((\alpha,\sigma)_{\alpha-1}, \sigma)_{\alpha}$, where we denote the fusion of a pair of anyons $\phi$ and $\phi'$ into a type $t$ anyon using the notation $(\phi,\phi')_t$.  We can represent these states using fusion trees or by depicting the anyon labels grouped by ovals labeling the total $q$-spin. 
\begin{equation}
\begin{split} \label{eq-single-subit}
 \ket{0} &= \hackcenter{\begin{tikzpicture}[scale=0.35]
          \draw[thick, fill=gray!30] (1.75,0,0) ellipse (3.4 and 1.45);
        \draw[thick,   fill=gray!30] (1,0,0) ellipse (2 and .7);
    \shade[ball color = red!40, opacity = 0.5] (0,0,0) circle (.35);
    \shade[ball color = red!40, opacity = 0.5] (2,0,0) circle (.35);
    \shade[ball color = red!40, opacity = 0.5] (4,0,0) circle (.35);
    \node at (0,0) {$\scriptstyle \alpha$};
    \node at (2,0) {$\scriptstyle \sigma$};
    \node at (4,0) {$\scriptstyle \sigma$};
            \node at (2.7,-.8) {$\scriptscriptstyle \alpha+1$};
            \node at (5,-1.15) {$\scriptstyle \alpha$};
\end{tikzpicture}} =
\hackcenter{\begin{tikzpicture}[ scale=1.1]
  \draw[ultra thick, black] (0,0) to (.6,-.6) to (.6,-1.2);
  \draw[ultra thick, black] (0.6,0) to (.3,-.3);
  \draw[ultra thick, black] (1.2,0) to (.6,-.6); 
   \node at (0,0.2) {$\alpha$};
   \node at (.6,0.2) {$\sigma$}; 
   \node at (1.2,0.2) {$\sigma$};  
    \node at (.8,-1) {$\scriptstyle \alpha$};
    \node at (.2,-.6) {$\scriptstyle \alpha + 1$}; 
\end{tikzpicture} }
\\
\ket{1} &= \hackcenter{\begin{tikzpicture}[scale=0.35]
          \draw[thick, fill=gray!30] (1.75,0,0) ellipse (3.4 and 1.45);
        \draw[thick,   fill=gray!30] (1,0,0) ellipse (2 and .7);
    \shade[ball color = red!40, opacity = 0.5] (0,0,0) circle (.35);
    \shade[ball color = red!40, opacity = 0.5] (2,0,0) circle (.35);
    \shade[ball color = red!40, opacity = 0.5] (4,0,0) circle (.35);
    \node at (0,0) {$\scriptstyle \alpha$};
    \node at (2,0) {$\scriptstyle \sigma$};
    \node at (4,0) {$\scriptstyle \sigma$};
            \node at (2.7,-.8) {$\scriptscriptstyle \alpha-1$};
            \node at (5,-1.15) {$\scriptstyle \alpha$};
\end{tikzpicture}} =
\hackcenter{\begin{tikzpicture}[ scale=1.1]
  \draw[ultra thick, black] (0,0) to (.6,-.6) to (.6,-1.2);
  \draw[ultra thick, black] (0.6,0) to (.3,-.3);
  \draw[ultra thick, black] (1.2,0) to (.6,-.6); 
   \node at (0,0.2) {$\alpha$};
   \node at (.6,0.2) {$\sigma$}; 
   \node at (1.2,0.2) {$\sigma$};  
    \node at (.8,-1) {$\scriptstyle \alpha$};
    \node at (.2,-.6) {$\scriptstyle \alpha - 1$}; 
\end{tikzpicture} }  
\end{split}
\end{equation}
Such fusion trees give rise to bases for the Hilbert space of a collection of anyons in a fixed $q$-spin.  An alternative single qubit Hilbert space is given by the states $\ket{0'} = (\alpha, (\sigma, \sigma)_{\1} )_{\alpha}$,
$\ket{1'} = ((\alpha,(\sigma, \sigma)_{\psi})_{\alpha}$.  The change of basis between these two bases is given by the matrix of $F$-symbols in the theory.  

For multiple qubits, we must account for the noncomputational fusion channels $\ket{NC_1} = ((\alpha,\sigma)_{\alpha+1}, \sigma)_{\alpha+2}$,
$\ket{NC_2} = ((\alpha,\sigma)_{\alpha-1}, \sigma)_{\alpha-2}$.  The computational basis of $n$ qubits is encoded into the fusion of $\alpha \times \sigma^{\times 2n}$ by having each 0 correspond to the fusion channel $\alpha \times \sigma \to \alpha +1$ and each 1 to $\alpha \times \sigma \to \alpha -1$ and each $(\alpha\pm 1) \times \sigma \to \alpha$ as illustrated below for $n=2$,  
\begin{align}  \label{eq:twoqubits}
& \hackcenter{\begin{tikzpicture}[scale=0.85]
  \draw[ultra thick, black] (0,0) to (1.2,-1.2);
  \draw[ultra thick, black] (0.6,0) to (.3,-.3);
  \draw[ultra thick, black] (1.2,0) to (.6,-.6);
  \draw[ultra thick, black] (1.8,0) to (.9,-.9);
  \draw[ultra thick, black] (2.4,0) to (1.2,-1.2) to (1.2,-1.8);
   \node at (0,0.2) {$\alpha$};
   \node at (.6,0.2) {$\sigma$};
   \node at (1.2,0.2) {$\sigma$};
   \node at (1.8,0.2) {$\sigma$};
   \node at (2.4,0.2) {$\sigma$};
   \node at (1.2,-2) {$\alpha$};
    \node at (.5,-.9) {$\scriptstyle a_2$};
    \node at (.18,-.6) {$\scriptstyle a_1$};
    \node at (.78,-1.2) {$\scriptstyle a_3$};
\end{tikzpicture}}
&\begin{aligned}
&\ket{00} = (\alpha+1, \alpha, \alpha+1) \\
&\ket{10} = (\alpha-1, \alpha, \alpha+1) \\
&\ket{01} = (\alpha+1, \alpha, \alpha-1) \\
&\ket{11} = (\alpha-1, \alpha, \alpha-1) \\
&\ket{NC_1} = (\alpha+1, \alpha+2, \alpha+1) \\
&\ket{NC_2} = (\alpha-1, \alpha-2, \alpha-1)
\end{aligned}  
\end{align}
where we encode the results of fusion as tuple $(a_1,a_2,a_3)$.

Note that while we use three anyons $(\alpha, \sigma, \sigma)$ to encode a single qubit, two qubits are encoded by adding just two additional Ising anyons $(\alpha, \sigma, \sigma,\sigma,\sigma)$.  No additional $\alpha$-type anyons are required in this multi qubit encoding.

Unlike the traditional framework of topological quantum computation, the non-semisimple theory gives rise to an inner product on the Hilbert space $\mathcal{H}_n$ that is not always positive definite~\cite{GLPMS}.  This is an inherent property of the theory. Nevertheless, there are sectors within the theory where the computational space is positive definite. 
For $\alpha \in (2,3)$, the inner product from \cite{GLPMS} is definite on the single qubit Hilbert space $\mathcal{H}_1$, but indefinite unitary more generally.  For example, the two qubit space $\mathcal{H}_2$ has signature $(+,+,+,+,-,+)$, meaning that all basis vectors from~(\ref{eq:twoqubits}) have norm $+1$ except $\braket{NC_1,NC_1} = -1$. Remarkably, the signature for the computational subspace is always positive definite for $\alpha \in (2,3)$ in the multi-qubit encoding~\footnote{Technically, we rescale the form from \cite{GLPMS}  by an overall sign.}, with indefinite signature only occurring on a portion of the noncomputational space. 
Braiding of anyons acts on the Hilbert space $\mathcal{H}_{n}$ by unitary transformations with respect to the indefinite metric.  In order to preserve the Hilbert space $\mathcal{H}_{\alpha;\sigma^n}$, we consider only those braids that take the $\alpha$-type anyon to itself.  This subset of braids is called the affine braid group and can be regarded as braids that are allowed to wrap around a stable, unmoved pole in the first strand.    
\[
\mathsf{b}_1^2 = \; 
\hackcenter{\begin{tikzpicture}[scale=0.35]
          \draw[thick, fill=gray!30] (1.75,0,0) ellipse (3.4 and 1.45);
        \draw[thick,   fill=gray!30] (1,0,0) ellipse (2 and .7);
    \shade[ball color = red!40, opacity = 0.5] (0,0,0) circle (.35);
    \shade[ball color = red!40, opacity = 0.5] (2,0,0) circle (.35);
    \shade[ball color = red!40, opacity = 0.5] (4,0,0) circle (.35);
    \node at (0,0) {$\scriptstyle \alpha$};
    \node at (2,0) {$\scriptstyle \sigma$};
    \node at (4,0) {$\scriptstyle \sigma$};
            \node at (2.7,-.8) {$\scriptscriptstyle \alpha\pm 1$};
            \node at (5,-1.15) {$\scriptstyle \alpha$};
    \draw[ultra thick, red,-] (0,3) to (0,5);  
                \path [ultra thick, fill=white] (-.2,3.5) rectangle (.2,4.1);
    \draw[ultra thick, black,-] (2,.5) to (2,.75).. controls ++(0,1.5) and ++(0,-1.2) ..   (-.75,3);
            \path[ultra thick, fill=white] (-.25,2) rectangle (.25,2.4);
\draw[ultra thick, black,-] (-.75,3) .. controls ++(0,1.2) and ++(0,-1.2) ..   (2,5);
    \draw[ultra thick, red,-] (0,.5) to (0,3);  
     \draw[ultra thick, black,-] (4,.5) to (4,5);
    \end{tikzpicture}}
    \quad 
    \mathsf{b}_2 = \;
\hackcenter{\begin{tikzpicture}[scale=0.35]
          \draw[thick, fill=gray!30] (1.75,0,0) ellipse (3.4 and 1.45);
        \draw[thick,   fill=gray!30] (1,0,0) ellipse (2 and .7);
    \shade[ball color = red!40, opacity = 0.5] (0,0,0) circle (.35);
    \shade[ball color = red!40, opacity = 0.5] (2,0,0) circle (.35);
    \shade[ball color = red!40, opacity = 0.5] (4,0,0) circle (.35);
    \node at (0,0) {$\scriptstyle \alpha$};
    \node at (2,0) {$\scriptstyle \sigma$};
    \node at (4,0) {$\scriptstyle \sigma$};
            \node at (2.7,-.8) {$\scriptscriptstyle \alpha\pm1$};
            \node at (5,-1.15) {$\scriptstyle \alpha$};
    \draw[ultra thick, red,-] (0,0.5) to (0,5);  
     \draw[ultra thick, black,-] (4,.5) to (4,2) .. controls ++(0,.8) and ++(0,-0.8) ..   (2,4) to (2,5) ;
    \path [fill=white] (2.8,1.6) rectangle (3.2,3.5);
\draw[ultra thick, black,-] (2,.5) to (2,2) .. controls ++(0,.8) and ++(0,-0.8) ..   (4,4) to (4,5);
    \end{tikzpicture}}
\]

Introduce real coefficients 
\begin{equation} \label{eq:Balphapmone}
   \mathbf{B}_{\alpha+1} := \cfrac{\sqrt{2}}{-1+\cot{\frac{\pi (\alpha+1)}{4}}},
 \;\; \;
   \mathbf{B}_{\alpha-1} := \cfrac{\sqrt{2}}{-1+\cot{\frac{\pi \alpha}{4}}},
\end{equation} 
that arise when computing the inner product on the computational Hilbert space (see Appendix~\ref{sec:inner}).  
The braiding matrices are given on the computational space by  
\begin{equation}\label{eq:single_qubit_braids}
\begin{split}
    \mathsf{b}_1^2 &= -q\left(\mathsf{b}_1^{\alpha \sigma \sigma}\right)^2 = \left(\begin{matrix}
        q^{\alpha} & 0\\
        0 & q^{-\alpha} 
    \end{matrix}\right),
\\
              \mathsf{b}_2 &= 
   q^{-\frac{3}{2}}\mathsf{b}_2^{\alpha\sigma\sigma}=q^{-1}\left(
 \begin{array}{cc}
  \frac{1+q^{2}}{1-q^{2\alpha}} & q^{-1}\frac{\sqrt{\mathbf{B}_{\alpha+1}}}{\sqrt{\mathbf{B}_{\alpha-1}}}\\
  q^{-1}\frac{\sqrt{\mathbf{B}_{\alpha+1}}}{\sqrt{\mathbf{B}_{\alpha-1}}} & \frac{1+q^{2}}{1-q^{-2\alpha}}
 \end{array}
 \right)
\end{split}
\end{equation}
where we sometimes include the superscript to emphasize this single qubit encoding. The factors $q^{-\frac{3}{2}}$ and $-q$ are global phases that were manually introduced to make the matrices special unitary.

An explicit computation shows that the matrices above satisfy the affine braid relation:
\begin{equation}\label{eq:affine_braid}
    \mathsf{b}_1^2 \mathsf{b}_2 \mathsf{b}_1^2 \mathsf{b}_2 = \mathsf{b}_2 \mathsf{b}_1^2 \mathsf{b}_2 \mathsf{b}_1^2
\end{equation}
More generally, we have elementary braid generators $\mathsf{b}_1^2$, $\mathsf{b}_2$, \dots, $\mathsf{b}_{2n}$ acting on the space $\mathcal{H}_n$ with $\mathsf{b}_i$ swapping $\sigma$'s in position $i$ and $(i+1)$ for $i>2$.

 \subsection{Single Qubit Operations} 

Universality for single qubit operations is established by showing that braiding will produce a dense subgroup of $SU(2)$. A standard approach from~\cite{Aharonov_2008}   
for showing density in $SU(2)$ is to specify two non-commuting braids $b_1, b_2$ of infinite order.  

If $\alpha \in \R \setminus \mathbb{Q}$, then $\mathsf{b}_1^2$ and $\mathsf{b}_2$  are non-commuting and $\mathsf{b}_1^2$ has infinite order, but $\mathsf{b}_2$ has order four.  However, $\mathsf{b}_2b_1 \mathsf{b}_2^{-1}$ will have infinite order.  Taking $b_1=\mathsf{b}_1^2$ and $b_2 = \mathsf{b}_2b_1 \mathsf{b}_2^{-1}$ gives two non-commuting matrices of infinite order;  hence, braiding anyons for any irrational $\alpha$ will be dense for single qubit operations.    

If $\alpha \in \Q$, it takes more work to prove density.   Here, we focus on the case $\alpha = 2 + \frac{2}{5}$, where we have an explicit procedure for constructing low-leakage two-qubit entangling gates. 
For rational $\alpha$, the braid $(\mathsf{b}_1^{\alpha \sigma \sigma})^2$ no longer has infinite order, so we instead let $b_1 = (\mathsf{b}_2^{\alpha \sigma \sigma }) (\mathsf{b}_1^{\alpha \sigma \sigma})^2(\mathsf{b}_2^{\alpha \sigma \sigma})^2$ and $b_2= \left(\mathsf{b}_2^{\alpha \sigma \sigma}\right) b_1 ({\mathsf{b}_2^{\alpha \sigma \sigma}})^{-1}$. $b_1$ and $b_2$ are not commuting, and they have infinite order if and only if $b_1$ does.
The eigenvalues of $b_1$ at $\alpha=2+\frac{2}{5}$ are $e^{\pm i\left(\arccos{-\frac{\varphi}{\sqrt{2}}} \right)}$, where $\varphi$ is the golden ratio, so this generator has infinite order as long as $\arccos{\left(-\frac{\varphi}{\sqrt{2}}\right)}$ is not a rational multiple of $\pi$. But all algebraic integers of order up to 4 that are twice the cosine of a rational multiple of $\pi$ are classified in \cite{PT_irrational_phase}. 
The order 4 algebraic integer $-\sqrt{2} \, \varphi$ (a root of $x^4-6 x^2+4$) is not included in the classified values.  Hence, braiding of anyons is universal for single qubit operations when $\alpha = 2 + \frac{2}{5}$.

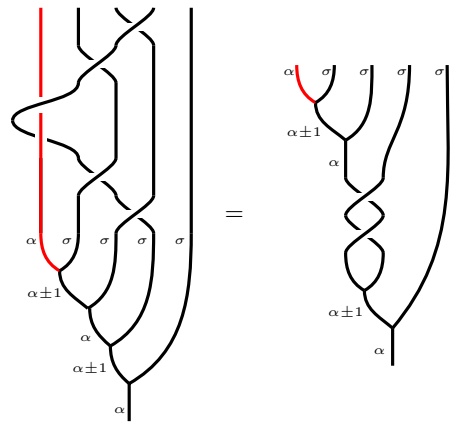
\begin{figure}
$\hackcenter{
\begin{tikzpicture}[  scale=0.5, yscale=-1  ]  
\draw [black, very thick]    (-.5,-2) .. controls +(0,.25) and +(0,-.25) ..  (.5,-1) ;
\path [fill=white] (-.25,-1.4) rectangle (.75,-1.6);
\draw [black, very thick]   (.5,-2)  .. controls +(0,.25) and +(0,-.25) ..  (-.5,-1);
\draw[very thick, ] (-1.5,-1) to (-1.5,-2);
\draw[very thick, ] (-2.5,-1) to (-2.5,-3);
\draw [black, very thick]    (-1.5,-3) .. controls +(0,.25) and +(0,-.25) ..  (-.5,-2) ;
\path [fill=white] (-1.25,-2.4) rectangle (-.25,-2.6);
\draw [black, very thick]   (-.5,-3)  .. controls +(0,.25) and +(0,-.25) ..  (-1.5,-2);
 \draw[very thick, ] (.5,-2) to (.5,-6);
 \draw [black, very thick]    (-3.25,-4) .. controls +(0,.45) and +(0,-.45) ..  (-1.5,-3) ;
 \path [fill=white] (-2.35,-3.35) rectangle (-2.65,-3.8);
  \draw [red, very thick]   (-2.5,-7) to  (-2.5,-1);
 \path [fill=white] (-2.35,-4.2) rectangle (-2.65,-4.6);
 \draw [black, very thick]   (-1.5,-5)  .. controls +(0,.5) and +(0,-.45) ..  (-3.25,-4);
 \draw[very thick, postaction={}] (-.5,-5) to (-.5,-3);
 \draw [black, very thick]    (-1.5,-6) .. controls +(0,.25) and +(0,-.25) ..  (-.5,-5) ;
 \path [fill=white] (-1.25,-5.4) rectangle (-.25,-5.6);
 \draw [black, very thick]   (-.5,-6)  .. controls +(0,.25) and +(0,-.25) ..  (-1.5,-5);
 \draw [black, very thick]    (-.5,-7) .. controls +(0,.25) and +(0,-.25) ..  (.5,-6) ;
 \path [fill=white] (-.25,-6.4) rectangle (.75,-6.6);
  \draw [black, very thick]   (.5,-7)  .. controls +(0,.25) and +(0,-.25) ..  (-.5,-6);
   \draw[very thick, ] (-1.5,-6) to (-1.5,-7);
   \draw[very thick,  ] (1.5,-1) to (1.5,-7);
\draw[very thick,] (-2,0) to [out=90, in=220] (-1.25,1);
\draw[very thick, ] (-.5,-1) to [out=90, in=-30] (-1.25,1);
\draw[very thick, ] (-1.2,1) to [out=90, in=220]  (-.65,2);
\draw[very thick,  red] (-2.5,-1)to [out=90, in=210] (-2,0);
\draw[very thick, ] (-1.5,-1) to [out=90, in=-30] (-2,0);
\draw[very thick,] (.5,-1) to [out=90, in=-30] (-.65,2);
\draw[very thick,] (-.65,2) to [out=90, in=220] (-.15,3);
\draw[very thick, ] (1.5,-1) to [out=90, in=-30] (-.15,3) to(-.15,4) ;
\node at (-2.75,-.8) {$\scriptscriptstyle  \alpha$};
\node at (-1.8,-.8) {$\scriptscriptstyle  \sigma$};
\node at (-0.8,-.8) {$\scriptscriptstyle  \sigma$};
\node at (.2, -.8) {$\scriptscriptstyle  \sigma$};
\node at (1.2,-.8) {$\scriptscriptstyle  \sigma$};
\node at (-1.3,1.8) {$\scriptscriptstyle \alpha $};
\node at (-2.4,.6) {$\scriptscriptstyle \alpha \pm 1$};
\node at (-1.2,2.6) {$\scriptscriptstyle \alpha \pm 1$};
\node at (-.4,3.7) {$\scriptscriptstyle  \alpha $};
\end{tikzpicture}}
\quad = \quad
 \hackcenter{
\begin{tikzpicture}[  scale=0.5,   yscale=-1 ]
\draw[very thick  ] (0,0) to [out=90, in=220] (.75,1);
\draw[very thick,  ] (2.2,-6) to [out=90, in=-50] (.75,1);
\draw[very thick, ] (.75,1) to  (.75,2);
\draw[very thick, ] (-.5,-1)to [out=90, in=210] (0,0);
\draw[very thick, ] (.5,-1) to [out=90, in=-30] (0,0);
\draw [black, very thick]    (-.5,-2) .. controls +(0,.25) and +(0,-.25) ..  (.5,-1) ;
\path [fill=white] (-.25,-1.4) rectangle (.75,-1.6);
\draw [black, very thick]   (.5,-2)  .. controls +(0,.25) and +(0,-.25) ..  (-.5,-1);
\draw [black, very thick]    (-.5,-3) .. controls +(0,.25) and +(0,-.25) ..  (.5,-2) ;
 \path [fill=white] (-.25,-2.4) rectangle (.75,-2.6);
 \draw [black, very thick]   (.5,-3)  .. controls +(0,.25) and +(0,-.25) ..  (-.5,-2);
\draw[very thick,] (-1.3,-5) to [out=90, in=220] (-.5,-4);
\draw[very thick, ] (.2,-6) to [out=90, in=-30] (-.5,-4);
\draw[very thick,  ] (-.5,-4) to  (-.5,-3);
\draw[very thick, red ] (-1.8,-6)to [out=90, in=210] (-1.3,-5);
\draw[very thick, ] (-.8,-6) to [out=90, in=-30] (-1.3,-5);
\draw[very thick, ] (1.2,-6) .. controls +(0,1.75) and +(0,-1) ..  (.5,-3);
\node at (-2,-5.8) {$\scriptscriptstyle  \alpha$};
\node at (-1,-5.8) {$\scriptscriptstyle  \sigma$};
\node at (-0,-5.8) {$\scriptscriptstyle  \sigma$};
\node at (1, -5.8) {$\scriptscriptstyle  \sigma$};
\node at (2,-5.8) {$\scriptscriptstyle  \sigma$};
\node at (-.8,-3.4) {$\scriptscriptstyle \alpha $};
\node at (-1.6,-4.2) {$\scriptscriptstyle \alpha \pm 1$};
\node at (-.5,0.6) {$\scriptscriptstyle \alpha \pm 1$};
\node at (.4,1.6) {$\scriptscriptstyle  \alpha $};
\end{tikzpicture} }$
    \caption{An illustration of the braid $J_4$ that acts diagonally on a single qubit away from the initial $\alpha$-type particle.  Because the topology of this braid is such that it can slide through the result of all previous fusions, this braid does nothing to qubits appearing earlier in the fusion tree.  Topologically, it is the same as a simple twist acting on a qubit lower in the fusion tree. }
    \label{fig:jucys_murphy}
\end{figure}

From our encoding of multiple qubits, it is not immediately obvious that braiding will produce single qubit operations on each qubit in the multi-qubit space since we use just a single $\alpha$-type anyon.   However, by leveraging special braids acting locally on a single qubit, the proof of universality for single qubit operations above immediately extends to the multiqubit encoding, see  Figure~\ref{fig:jucys_murphy}. 

In the basis of the two qubit space $\mathcal{H}_2$ for $\alpha \times \sigma^{\times 4}$ from \eqref{eq:twoqubits}, single qubit operations can be performed on the first qubit using the affine braid generators \begin{equation}\label{eq:multi_SU2_density}
\begin{split}
    \mathsf{b}_1^{2}   &= \left(\mathsf{b}_1^{\alpha \sigma\sigma }\right)^2 \otimes \Id \oplus \left(\mathsf{b}_1^{\alpha \sigma\sigma}\right)^2
\\
    \mathsf{b}_2  &= \mathsf{b}_2^{\alpha  \sigma\sigma} \otimes \Id \oplus \left(q^\frac{1}{2} \Id \right)
    \end{split}
\end{equation}
Where $\mathsf{b}^{\alpha \sigma\sigma}$ refers to the matrices in~(\ref{eq:single_qubit_braids}). 
The single qubit operations on the second qubit can be achieved using 
\begin{equation}\label{eq:multi_SU2_density_second_qubit}
\begin{split}
    J_4 &= \Id \otimes \left(\mathsf{b}_1^{\alpha \sigma\sigma}\right)^2 \oplus \left(\begin{matrix}
        q^{1-\alpha} & 0\\
        0 & q^{1+\alpha}
    \end{matrix} \right)
\\
    \mathsf{b}_4  &= \Id \otimes  \mathsf{b}_2^{\alpha \sigma\sigma}\oplus \left(q^\frac{1}{2} \Id \right)
    \end{split}
\end{equation}
where $J_4$ is the composite braid $\mathsf{b}_3\mathsf{b}_2\mathsf{b}_1^2\mathsf{b}_2 \mathsf{b}_3$.
The first four basis vectors from \eqref{eq:twoqubits} span the computational space, so $J_4$ and $b_4$ introduce no leakage in executing single-qubit operations. 
 
The structure of~(\ref{eq:multi_SU2_density}) and~(\ref{eq:multi_SU2_density_second_qubit}) simplifies the compilation of single-qubit gates in the multi-qubit setting; knowing how to implement a gate on the first qubit automatically gives us a way of implementing it on the second qubit. For example, if a sequence of powers of $(\mathsf{b}_1^2)$ and $\mathsf{b}_2$ generates the gate $U\otimes \Id$, $U\in SU(2)$, then the same sequence in $J_4$ and $\mathsf{b}_4$ will generate the gate $\Id\otimes U$.  

\subsection{Entangling Gates} 
Given the ability to efficiently perform single-qubit operations on any qubit, achieving universal quantum computation also requires the implementation of entangling gates and controlling the leakage into the non-computational subspace. In many approaches to topological quantum computation, density is established across the entire anyonic Hilbert space, which implies density on the computational subspace. However, the challenge then shifts to constructing entangling gates that minimize leakage into non-computational states.

In our setting, it is more challenging to establish density on the full Hilbert space. In the two qubit setting, the 6-dimensional Hilbert space $\mathcal{H}_2$ has signature $(+,+,+,+,-,+)$ and braiding produces a subgroup of $SU(5,1)$.  Unlike the compact group $SU(6)$, this group of matrices preserving an indefinite form is noncompact and the classification of its subgroups is much more complex.   Very few techniques have been established for proving density in this setting~\cite{Aharonov_2011,GLPMS4, Kupdense}. 

Rather than proving density for the entire Hilbert space $\mathcal{H}_n$, here we establish density only for the computational space.  Our approach combines two techniques developed in the context of Fibonacci anyons. The first is a technique to apply controlled braiding where the state of a control qubit determines if a braid is executed on a target qubit following ideas from \cite{PhysRevB.75.165310,PhysRevLett.96.070503,PhysRevLett.95.140503}. 
Second, we have a simple and efficient procedure for finding specific braids that produce diagonal phases with arbitrarily small leakage into the noncomputational space, adapting a procedure developed by Reichardt~\cite{Reichdist} to the indefinite unitary setting.  Together, these two strategies produce controlled phase gates.

\begin{figure}[htp!]\label{fig:alternative_basis}  
    \centering
$
 \hackcenter{\begin{tikzpicture}[scale=.8]
  \draw[ultra thick, black] (0,0) to (1.2,-1.2);
  \draw[ultra thick, black] (0.6,0) to (.9,-.3);
  \draw[ultra thick, black] (1.2,0) to (.9,-.3);
        \draw[ultra thick, black, dotted] (0.6,-.6) to (.9,-.3);
  \draw[ultra thick, black] (1.8,0) to (.9,-.9);
  \draw[ultra thick, black] (2.4,0) to (1.2,-1.2) to (1.2,-1.8);
   \node at (0,0.2) {$\alpha$};
   \node at (.6,0.2) {$\sigma$}; 
   \node at (1.2,0.2) {$\sigma$}; 
   \node at (1.8,0.2) {$\sigma$}; 
   \node at (2.4,0.2) {$\sigma$}; 
   \node at (1.2,-2) {$\alpha$};
    \node at (.62,-.9) {$\scriptstyle \alpha$};
   \node at (.95,-.5) {$\scriptstyle \1$};
    \node at (.75,-1.2) {$\scriptstyle \alpha+ 1$};
\end{tikzpicture} }
\quad 
 \hackcenter{\begin{tikzpicture}[scale=.8]
  \draw[ultra thick, black] (0,0) to (1.2,-1.2);
  \draw[ultra thick, black] (0.6,0) to (.9,-.3);
  \draw[ultra thick, black] (1.2,0) to (.9,-.3);
        \draw[ultra thick, black, dotted] (0.6,-.6) to (.9,-.3);
  \draw[ultra thick, black] (1.8,0) to (.9,-.9);
  \draw[ultra thick, black] (2.4,0) to (1.2,-1.2) to (1.2,-1.8);
   \node at (0,0.2) {$\alpha$};
   \node at (.6,0.2) {$\sigma$}; 
   \node at (1.2,0.2) {$\sigma$}; 
   \node at (1.8,0.2) {$\sigma$}; 
   \node at (2.4,0.2) {$\sigma$}; 
   \node at (1.2,-2) {$\alpha$};
    \node at (.62,-.9) {$\scriptstyle \alpha$};
   \node at (.95,-.5) {$\scriptstyle \1$};
    \node at (.75,-1.2) {$\scriptstyle \alpha -1$};
\end{tikzpicture} }
\quad 
    \hackcenter{\begin{tikzpicture}[scale=.8]
  \draw[ultra thick, black] (0,0) to (1.2,-1.2);
  \draw[ultra thick, black] (0.6,0) to (.9,-.3);
  \draw[ultra thick, black] (1.2,0) to (.6,-.6);
  \draw[ultra thick, black] (1.8,0) to (.9,-.9);
  \draw[ultra thick, black] (2.4,0) to (1.2,-1.2) to (1.2,-1.8);
   \node at (0,0.2) {$\alpha$};
   \node at (.6,0.2) {$\sigma$}; 
   \node at (1.2,0.2) {$\sigma$}; 
   \node at (1.8,0.2) {$\sigma$}; 
   \node at (2.4,0.2) {$\sigma$}; 
   \node at (1.2,-2) {$\alpha$};
    \node at (.45,-.9) {$\scriptstyle \alpha +2 $};
    \node at (.7,-1.2) {$\scriptstyle \alpha +1$};
       \node at (.95,-.5) {$\scriptstyle \psi$};
\end{tikzpicture} }
$  $
    \hackcenter{\begin{tikzpicture}[scale=.8]
  \draw[ultra thick, black] (0,0) to (1.2,-1.2);
  \draw[ultra thick, black] (0.6,0) to (.9,-.3);
  \draw[ultra thick, black] (1.2,0) to (.6,-.6);
  \draw[ultra thick, black] (1.8,0) to (.9,-.9);
  \draw[ultra thick, black] (2.4,0) to (1.2,-1.2) to (1.2,-1.8);
   \node at (0,0.2) {$\alpha$};
   \node at (.6,0.2) {$\sigma$}; 
   \node at (1.2,0.2) {$\sigma$}; 
   \node at (1.8,0.2) {$\sigma$}; 
   \node at (2.4,0.2) {$\sigma$}; 
   \node at (1.2,-2) {$\alpha$};
    \node at (.62,-.9) {$\scriptstyle \alpha$};
    \node at (.75,-1.2) {$\scriptstyle \alpha + 1$};
       \node at (.95,-.5) {$\scriptstyle \psi$};
\end{tikzpicture} }
\quad 
    \hackcenter{\begin{tikzpicture}[scale=.8]
  \draw[ultra thick, black] (0,0) to (1.2,-1.2);
  \draw[ultra thick, black] (0.6,0) to (.9,-.3);
  \draw[ultra thick, black] (1.2,0) to (.6,-.6);
  \draw[ultra thick, black] (1.8,0) to (.9,-.9);
  \draw[ultra thick, black] (2.4,0) to (1.2,-1.2) to (1.2,-1.8);
   \node at (0,0.2) {$\alpha$};
   \node at (.6,0.2) {$\sigma$}; 
   \node at (1.2,0.2) {$\sigma$}; 
   \node at (1.8,0.2) {$\sigma$}; 
   \node at (2.4,0.2) {$\sigma$}; 
   \node at (1.2,-2) {$\alpha$};
    \node at (.62,-.9) {$\scriptstyle \alpha$};
    \node at (.75,-1.2) {$\scriptstyle \alpha- 1$};
       \node at (.95,-.5) {$\scriptstyle \psi$};
\end{tikzpicture} }
\quad 
    \hackcenter{\begin{tikzpicture}[scale=.8]
  \draw[ultra thick, black] (0,0) to (1.2,-1.2);
  \draw[ultra thick, black] (0.6,0) to (.9,-.3);
  \draw[ultra thick, black] (1.2,0) to (.6,-.6);
  \draw[ultra thick, black] (1.8,0) to (.9,-.9);
  \draw[ultra thick, black] (2.4,0) to (1.2,-1.2) to (1.2,-1.8);
   \node at (0,0.2) {$\alpha$};
   \node at (.6,0.2) {$\sigma$}; 
   \node at (1.2,0.2) {$\sigma$}; 
   \node at (1.8,0.2) {$\sigma$}; 
   \node at (2.4,0.2) {$\sigma$}; 
   \node at (1.2,-2) {$\alpha$};
    \node at (.45,-.9) {$\scriptstyle \alpha -2 $};
    \node at (.7,-1.2) {$\scriptstyle \alpha- 1$};
       \node at (.95,-.5) {$\scriptstyle \psi$};
\end{tikzpicture} }
$
  \caption{An alternative basis for the two-qubit space $\mathcal{H}_2$ that is well suited for performing controlled operations.  The first two $\sigma$-type anyons form the control pair.  Braiding the pair through the other anyon types creates a controlled braiding operation that only acts nontrivially if the control pair fuses into the $\psi$ channel. The subspace $\mathcal{H}_2^{\psi}$ consisting of states where the control qubit is in state $\psi$ has basis given by the last four basis vectors that we denote by $\ket{\psi_0}$, $\ket{\psi_1}$, $\ket{\psi_2}$, $\ket{\psi_3}$. 
  }  \label{img:controlled_gate_basis}  
\end{figure}
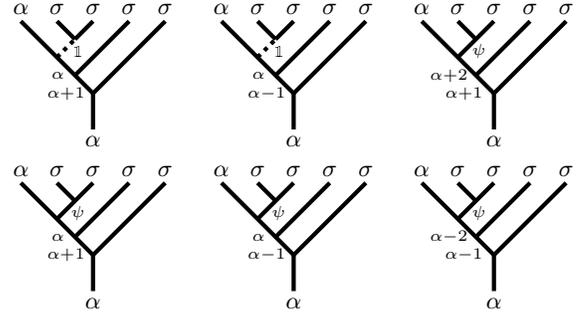

 The protocol for implementing controlled braid operations is built off the observation that $\sigma \times \sigma$ can fuse into the vacuum $\1$, or a $\psi$ type anyon.   Thus, braiding a pair of $\sigma$ anyons through other anyons will produce a trivial operation when they are in the state $\1$, and a nontrivial operation when in state $\psi$. 
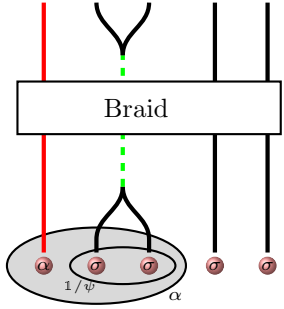
 \[ 
\hackcenter{\begin{tikzpicture}[scale=0.35]
          \draw[thick, fill=gray!30] (2,0,0) ellipse (3.4 and 1.45);
        \draw[thick,   fill=gray!30] (3,0,0) ellipse (2 and .7);
    \shade[ball color = red!40, opacity = 0.5] (0,0,0) circle (.35);
    \shade[ball color = red!40, opacity = 0.5] (2,0,0) circle (.35);
    \shade[ball color = red!40, opacity = 0.5] (4,0,0) circle (.35);
    \shade[ball color = red!40, opacity = 0.5] (6.5,0,0) circle (.35);
    \shade[ball color = red!40, opacity = 0.5] (8.5,0,0) circle (.35);
    \node at (0,0) {$\scriptstyle \alpha$};
    \node at (2,0) {$\scriptstyle \sigma$};
    \node at (4,0) {$\scriptstyle \sigma$};
    \node at (6.5,0) {$\scriptstyle \sigma$};
    \node at (8.5,0) {$\scriptstyle \sigma$};
            \node at (1.35,-.8) {$\scriptscriptstyle \1/\psi$};
            \node at (5,-1.15) {$\scriptstyle \alpha$};
    \draw[ultra thick, red,-] (0,0.5) to (0,10);  
    \draw[ultra thick, black,-] (6.5,0.5) to (6.5,10);   
    \draw[ultra thick, black,-] (8.5,0.5) to (8.5,10); 
   \draw[ultra thick, black,-] (2,.5) to (2,1) to[out=90,in=-90] (3,3);
   \draw[ultra thick, black,-] (4,.5) to (4,1) to[out=90,in=-90] (3,3);
   \draw[ultra thick, green,dashed] (3,3) to   (3,8);
   \draw [thick , fill=white ] (-1,5) rectangle (9,7);
    \draw[ultra thick, black,-]   (2,10) to[out=-90,in=90] (3,8);
   \draw[ultra thick, black,-]  (4,10) to[out=-90,in=90] (3,8);
   \node at (3.5,6) {Braid };
    \end{tikzpicture}}
\]
 Changing basis in the control qubit to the basis $\ket{0'} = (\alpha, (\sigma, \sigma)_{\1} )_{\alpha}$,
$\ket{1'} = ((\alpha,(\sigma, \sigma)_{\psi})_{\alpha}$ (see Figure~\ref{img:controlled_gate_basis}), we give braids that act as the identity when the control channel is $\1$ and nontrivially when it is
$\psi$. The gate implemented by this protocol is, by definition, a controlled gate in this basis.  Transforming back to the original basis maintains the entanglement.

For controlled braiding to entangle the computational space, we must identify low-leakage braids acting on the Hilbert space $\mathcal{H}_2^{\psi}$ corresponding to the anyon configuration $(\alpha, \psi, \sigma^{2} )$ in a total $q$-spin state $\alpha$.  We focus on braids that preserve this space and become topologically trivial when $\psi$ is replaced by the vacuum $\1$.  Fix a basis of $\mathcal{H}_2^{\psi}$ consisting of the last four basis elements in Figure~\ref{img:controlled_gate_basis} containing the fusion of $\sigma \sigma$ into a $\psi$ state. 
We denote these by $\ket{\psi_0}$, $\ket{\psi_1}$, $\ket{\psi_2}$ $\ket{\psi_3}$, noting that $\ket{\psi_0}$ and $\ket{\psi_3}$ span the noncomputational space.   

Our strategy to obtain low-leakage gate acting on $\mathcal{H}_2^{\psi}$ is to start with a gate $w=U\oplus V$ in $SU(2) \times SU(1,1)$ and perform a sequence of braids to suppress the off-diagonal entries so that the resulting gate approximates a diagonal with a nontrivial phase on the computational subspace.  Here $SU(2)$ acts on the space spanned by $\ket{\psi_0}$, $\ket{\psi_1}$ and $SU(1,1)$ acts on the space spanned by $\ket{\psi_2}$ and $\ket{\psi_3}$, where $\ket{\psi_1}$ and $\ket{\psi_2}$ span the computational space.
 
For matrices  $U \in SU(2)$, Reichardt gave a procedure for leveraging a specific diagonal matrix~\footnote{Reichardt used a slightly different definition of $D_{(2)}$.   }
$D_{(2)} = \left( \begin{matrix}
    e^{\frac{i3\pi}{5}} & 0\\
    0 & 1\\
\end{matrix}\right)$
to produce a sequence of unitaries $U_k$ whose off-diagonal entries decrease exponentially~\cite[Equation 3]{Reichdist}.  The recursive sequence is given as follows:
\begin{equation} \label{eq:reichardt_unitary}
    \begin{split}
         U_0 &= U \\
    U_{k+1} &= U_k D_{(2)}U_k^{\dag}  D_{(2)}^3 U_k D_{(2)}^3 U_k^\dag D_{(2)} U_k
    \end{split}
\end{equation}
One computes that the off diagonal entries satisfy $|\langle 0 | U_{k+1} | 1 \rangle| = |\langle 0 | U_k | 1 \rangle|^5$, and hence that $|\langle 0 | U_{k} | 1 \rangle| = |\langle 0 | U | 1 \rangle|^{5^k}$. For unitary matrices, the off-diagonal entries are less than or equal to 1, so iterating this procedure suppresses the off-diagonal entries to arbitrary accuracy.  Note that, even though the sequence in~(\ref{eq:reichardt_unitary}) has an exponential length in $k$, the off-diagonal terms go to zero doubly exponentially.  While it is not possible to control the phases of the resulting diagonal matrix, one can still use the algorithm to produce entangling gates \cite{carnahan_systematically_generate}.

Recall that matrices $V$ in $SU(1,1)$ have unit determinant and satisfy $V^{\dag} J V= J$ where $J = \left(
  \begin{array}{cc}
    1 & 0 \\
   0 & -1 \\
  \end{array}
\right)$.  The general form of such a matrix is 
$V=  \left(\begin{array}{cc}
    \alpha & \beta \\
   \beta^\ast & \alpha^{\ast} \\
  \end{array}\right)$
where $|\alpha|^2 - |\beta|^2 =1$.   Unlike the case of $SU(2)$, the off-diagonal entries $\beta$ need not have a norm less than or equal to 1, so Reichardt's algorithm does not immediately apply.   However, for  
 $V \in SU(1,1)$ and 
$D_{(1, 1)} =  \left( \begin{matrix}
    1 & 0\\
    0 & e^{\frac{i3\pi}{5}}\\
\end{matrix}\right)$,  and setting 
\begin{equation} \label{eq:reichardt_indefinite_unitary}
\begin{split}
    V_0 &= V \\ 
    V_{k+1} &= V_k D_{(1, 1)} V_k^{-1} D_{(1,  1)}^3 V_k D_{(1, 1)}^3 V_k^{-1} D_{(1, 1)} V_k
\end{split}
\end{equation} 
a direct computation shows that $|\langle 0 | V_{k+1} | 1 \rangle| = |\langle 0 | V_k | 1 \rangle|^5$.  Thus, if $V$ can be found with $|\langle 0 | V | 1 \rangle| <1$, Reichardt's algorithm will produce a sequence of matrices in $SU(1,1)$ with exponentially suppressed off-diagonal terms. 

Combining these two constructions, for $D=D_{(2)} \oplus D_{(1,1)}$, and $W=U \oplus V \in SU(2) \oplus SU(1,1)$, the recursive sequence 
\begin{equation} \label{eq:mixed-Reichardt}
\begin{split}
    W_0 &= W \\ 
    W_{k+1} &= W_k D  W_k^{-1} D^3 W_k D^3 W_k^{-1} D  W_k
\end{split}
\end{equation}
then in the ordered basis $\{ \ket{\psi_0}, \ket{\psi_1}, \ket{\psi_2}, \ket{\psi_3}\}$ we have  $|\langle \psi_0 | W_{k} | \psi_1 \rangle| = |\langle \psi_0 | W | \psi_1 \rangle|^{5^k}$ and $|\langle \psi_2 | W_{k} | \psi_3 \rangle| = |\langle \psi_2 | W | \psi_3 \rangle|^{5^k}$.  In particular, if $|\langle \psi_0 | W | \psi_1 \rangle| < 1$ and $|\langle \psi_2 | W | \psi_3 \rangle|<1$, this recursive sequence produces reduced leakage into the noncomputational space.  

\begin{figure}
    \centering
    $ W = \hackcenter{\begin{tikzpicture}[  scale=0.5 ]
    \draw[black, very thick]        (2, -5) to (2, 5);
    \begin{scope}[shift={(0,5.2)}]
        \draw[red, very thick]      (-1, 0) to (-1,-0.2);
        \draw[green, very thick]    ( 0, 0) to ( 0,-0.2);
        \draw[black, very thick]    ( 1, 0) to ( 1,-0.2);
        \draw[black, very thick]    ( 2, 0) to ( 2,-0.2);
    \end{scope}
    \begin{scope}[shift={(0,4)}]
        \draw [black, very thick]   (1.0,0) to  (1.0,1);
        \path [fill=white]          (0.85,.65) rectangle (1.15, .2);
        \draw [green, very thick]   (1.75,0)  .. controls +(0,.5) and +(0,-.45) ..  (0,1);
        \draw [red, very thick]     (-1.0,0) to  (-1.0,1);
    \end{scope}
    \begin{scope}[shift={(0, 2)}]
        \draw [green, very thick]   (-1.75,0)  .. controls +(0,.5) and +(0,-.45) ..  (1.75,2);        
        \path [fill=white]          (0.85, 0.4+.85) rectangle (1.15, 0.8+.85);
        \draw [black, very thick]   (1.0,0) to  (1.0,2);
        \path [fill=white]          (0.85-2, 0.4) rectangle (1.15-2, 0.8);
        \draw [red, very thick]     (-1.0,0) to  (-1.0,2);
    \end{scope}
    \begin{scope}[shift={(0,0)}]
        \draw [red, very thick]     (-1, 0) to (-1, 2);
        \path [fill=white]          (0.85-2, 0.4+.85) rectangle (1.15-2, 0.8+.85);
        \draw [green, very thick]   (1.75,0)  .. controls +(0,.5) and +(0,-.45) ..  (-1.75,2);
        \path [fill=white]          (0.85, 0.4) rectangle (1.15, 0.8);
        \draw [black, very thick]   ( 1, 0) to ( 1, 2);
    \end{scope}
    \begin{scope}[shift={(0,-2)}]
        \draw [black, very thick]   (1.0,0) to  (1.0,2);
        \path [fill=white]          (0.85, 0.4+.85) rectangle (1.15, 0.8+.85);
        \draw [green, very thick]   (-1.75,0)  .. controls +(0,.5) and +(0,-.45) ..  (1.75,2);
        \path [fill=white]          (0.85-2, 0.4) rectangle (1.15-2, 0.8);
        \draw [red, very thick]     (-1.0,0) to  (-1.0,2);
    \end{scope}
    \begin{scope}[shift={(0,-4)}]
        \draw [red, very thick]     (-1, 0) to (-1, 2);
        \path [fill=white]          (0.85-2, 0.4+.85) rectangle (1.15-2, 0.8+.85);
        \draw [green, very thick]   (1.75,0)  .. controls +(0,.5) and +(0,-.45) ..  (-1.75,2);
        \path [fill=white]          (0.85, 0.4) rectangle (1.15, 0.8);
        \draw [black, very thick]   ( 1, 0) to ( 1, 2);
    \end{scope}
    \begin{scope}[shift={(0,-5)}]
        \draw [black, very thick]   (1.0,0) to  (1.0,1);
        \path [fill=white]          (0.85,0.4) rectangle (1.15,0.8);
        \draw [green, very thick]   (0,0)  .. controls +(0,.5) and +(0,-.45) ..  (1.75,1);
        \draw [red, very thick]     (-1.0,0) to  (-1.0,1);
    \end{scope}
    \begin{scope}[shift={(0,-5.2)}]
        \draw[red, very thick]      (-1, 0) to (-1, 0.2);
        \draw[green, very thick]    ( 0, 0) to ( 0, 0.2);
        \draw[black, very thick]    ( 1, 0) to ( 1, 0.2);
        \draw[black, very thick]    ( 2, 0) to ( 2, 0.2);
    \end{scope}
    \begin{scope}[shift={(0,-5.6)}]
        \node at (-1, 0) {$\alpha$};
        \node at ( 0, 0) {$\psi$};
        \node at ( 1, 0) {$\sigma$};
        \node at ( 2, 0) {$\sigma$};
    \end{scope}
\end{tikzpicture}} \quad D =  \hackcenter{\begin{tikzpicture}[  scale=0.5 ]
    \draw[black, very thick]        (2, -5) to (2, 5);
    \begin{scope}[shift={(0,5.2)}]
        \draw[red, very thick]      (-1, 0) to (-1,-3.2);
        \draw[green, very thick]    ( 0, 0) to ( 0,-3.2);
        \draw[black, very thick]    ( 1, 0) to ( 1,-3.2);
        \draw[black, very thick]    ( 2, 0) to ( 2,-3.2);
    \end{scope}
    \begin{scope}[shift={(0,1)}]
        \draw [red, very thick]     (-1.0,0) to  (-1.0,1);
        \path [fill=white]          (-0.85,.65) rectangle (-1.15, .2);
        \draw [green, very thick]   (-1.75,0)  .. controls +(0,.5) and +(0,-.45) ..  (0,1);
        \draw [black, very thick]   (1.0,0) to  (1.0,1);
    \end{scope}
    \begin{scope}[shift={(0,0)}]
        \draw [green, very thick]   (0,0)  .. controls +(0,.5) and +(0,-.45) ..  (-1.75,1);
        \path [fill=white]          (-0.85,0.4) rectangle (-1.15,0.8);
        \draw [black, very thick]   (1.0,0) to  (1.0,1);
        \draw [red, very thick]     (-1.0,0) to  (-1.0,1);
    \end{scope}
    \begin{scope}[shift={(0,-1)}]
        \draw [red, very thick]     (-1.0,0) to  (-1.0,1);
        \path [fill=white]          (-0.85,.65) rectangle (-1.15, .2);
        \draw [green, very thick]   (-1.75,0)  .. controls +(0,.5) and +(0,-.45) ..  (0,1);
        \draw [black, very thick]   (1.0,0) to  (1.0,1);
    \end{scope}
    \begin{scope}[shift={(0,-2)}]
        \draw [green, very thick]   (0,0)  .. controls +(0,.5) and +(0,-.45) ..  (-1.75,1);
        \path [fill=white]          (-0.85,0.4) rectangle (-1.15,0.8);
        \draw [black, very thick]   (1.0,0) to  (1.0,1);
        \draw [red, very thick]     (-1.0,0) to  (-1.0,1);
    \end{scope}
    \begin{scope}[shift={(0,-5.2)}]
        \draw[red, very thick]      (-1, 0) to (-1, 3.2);
        \draw[green, very thick]    ( 0, 0) to ( 0, 3.2);
        \draw[black, very thick]    ( 1, 0) to ( 1, 3.2);
        \draw[black, very thick]    ( 2, 0) to ( 2, 3.2);
    \end{scope}
    \begin{scope}[shift={(0,-5.6)}]
        \node at (-1, 0) {$\alpha$};
        \node at ( 0, 0) {$\psi$};
        \node at ( 1, 0) {$\sigma$};
        \node at ( 2, 0) {$\sigma$};
    \end{scope}
\end{tikzpicture}}
$
    \caption{The braids $W$ and $D$. Note that if the control strand (green) is the vacuum, the braids are topologically trivial.}
    \label{img:w_braid}
\end{figure}
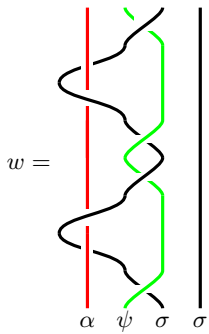

This is where our specific choice $\alpha = 2 +2/5$  enters our analysis.  For this value of $\alpha$, a direct computation shows that $(\mathsf{b}_1^{\alpha \psi \sigma \sigma })^4 = D_{(2)} \oplus D_{(1, 1)}$.   It remains to identify a braid $W$ of the form $W=U \oplus V$ that is topologically trivial when the strand labeled $\psi$ is removed and has small leakage term $|\langle \psi_2 | W | \psi_3 \rangle|<1$. One candidate braid is $\left(\mathsf{b}_2^{\alpha \psi \sigma \sigma} \right)^2$ but when $\alpha = 2+2/5$, $\left(\mathsf{b}_2^{\alpha \psi \sigma \sigma} \right)^2$ has non unitary leakage $|\langle \psi_0 | W | \psi_1 \rangle| \sim 1.943>1$. Another candidate is the 
restriction of the braid $W:=\left(\mathsf{b}_2^{\alpha \psi \sigma \sigma} \right)^2\left(\mathsf{b}_1^{\alpha \psi \sigma \sigma} \right)^2\left(\mathsf{b}_2^{\alpha \psi \sigma \sigma} \right)^2\left(\mathsf{b}_1^{\alpha \psi \sigma \sigma} \right)^2\left(\mathsf{b}_2^{\alpha \psi \sigma \sigma} \right)^{-2}$ from Figure~\ref{img:w_braid} to the four-dimensional space $\mathcal{H}_2^{\psi}$; it has the form $W= U \oplus V$, where $U\ \in SU(2)$ and $V \in SU(1,1)$ and acts as the identity if the control channel is in the vacuum state.

$W$ has leakage terms $|\langle \psi_0 | W | \psi_1 \rangle| \sim 0.832<1$ and $|\langle \psi_2 | W | \psi_3 \rangle| \sim 0.904<1$, ensuring the convergence of Reichardt's procedure. When said procedure is applied once, the non-unitary leakage term's norm is brought down to $\sim 0.399$. A second iteration reduces it to $\sim 0.101$. Braids with lower overhead can be found through a bruteforce search: \begin{equation*}
\mathsf{b}_2\mathsf{b}_1\mathsf{b}_2\mathsf{b}_1^{-1}\mathsf{b}_2^{-1}\mathsf{b}_1\mathsf{b}_2\mathsf{b}_1\mathsf{b}_2\mathsf{b}_1^2\mathsf{b}_2
\end{equation*}
has initial unitary and non-unitary leakage terms with norms $\sim 0.286$ and $\sim 0.285$ respectively; these norms both reduce to  $\sim 1.914 \times 10^{-3}$ upon applying Reichardt's prescription once, and to $2.565\times 10^{-14}$ after a second application.
This modification of Reichardt's algorithm produces an approximate diagonal, non-leaking gate. We just have to verify that it doesn't act like (a multiple of) the identity on the computational space. This is easy to verify numerically; analyzing the third iteration of Reichardt's sequence \eqref{eq:mixed-Reichardt} with starting braid $W$, we see that the diagonal gate acts on the computational space as diag$(e^{i\theta_1}, e^{i \theta_2})$ where $\theta_1 \sim -1.772 $ and $\theta_2  \sim -1.682$. The doubly exponential convergence rate ensures that further iterations will not meaningfully change these values \cite{carnahan_systematically_generate}.

\section{Discussion}
We have demonstrated that the new theories of non-semisimple TQFTs can offer more powerful models for topological quantum computation compared to traditional approaches based on semisimple theories. Specifically, we showed that the non-semisimple $SU(2)_2$ theory can serve as a universal model for quantum computation through braiding alone, whereas the more familiar semisimple $SU(2)_2$ theory of Ising anyons is not capable of achieving universality through braiding alone.

The passage from semisimple to non-semisimple TQFTs presents new challenges associated with indefinite unitarity of the inner products on the anyonic Hilber spaces.   We demonstrated that this obstacle can be overcome by producing an anyon encoding where the computational Hilbert space has a positive definite inner product and the indefiniteness of the inner product is constrained to the noncomputational space.   Further, we adapted an algorithm of Reichardt to produce a simple iterative approach to producing low leakage entangling gates.

As with any theory of topological quantum computation, its physical relevance depends on identifying physical models that support nonabelian quasiparticles. We have argued that, much like the promising outlook for traditional $SU(2)_k$ anyon theories within fractional quantum Hall systems, our approach is similarly motivated by the success of conformal field theories in providing effective descriptions of such systems. The theories presented here have a close connection to \textit{logarithmic} conformal field theories and show potential for realization within condensed matter systems, offering a pathway for discovering new, physically viable platforms for topological quantum computation already at the $\nu=5/2$ filling fraction.

In future work, we plan to extend the values of $\alpha$ that produce universal computation.  Producing low-leakage entangling gates will require new techniques, as Reichardt's algorithm does not readily generalize.   It is possible that some variant of the injection weave from \cite{PhysRevLett.96.070503,PhysRevLett.95.140503} can be adapted.   Further, it will be advantageous to develop new methods for proving density in the indefinite unitary group to establish density on the full anyonic Hilbert space $\mathcal{H}_n$, rather than just the protected computational subspace.  As it is likely that the full Hilbert space of such theories will be indefinite unitary, developing an analogous variant of the Solvey-Kitaev algorithm for noncompact groups $SU(n,m)$ will also be important.   

Finally, here we used the mapping class group of points on a disc to construct our computational model.  To address the question of anyons on more general surfaces, one would need to adapt the spin 3-manifold invariant in \cite{spin} to a full TQFT. 

\section{Methods} \label{sec:Appendix}
This section compiles the numerical data of our states and unitary braid operators. These are encoded using the categorical data organized in a general framework called \textit{non-semisimple modular categories}. A primary example is the category of finite-dimensional representations over the restricted unrolled quantum group with the quantum parameter set to a root of unity. Our framework focuses on this category for the case of $\mf{sl}(2,\C)$ with the quantum parameter specialized at an eighth root of unity.

\subsection{Modified Quantum Dimensions}
In the traditional approach to topological quantum computing, representations with quantum dimension zero, or {\em negligible} representations, are removed to form a semisimple category with a finite number of anyon types.   The new non-semisimple technology leverages these previously `negligible' representations to form new anyon types, the $\alpha$-type anyons studied here~\footnote{Given that the `negligible' representations now play a prominent role in non-semisimple TQFT and our construction of universal topological quantum computation, these previously neglected new anyon types might appropriately be called {\em neglectons}.}. 
The main technology used to resurrect the $\alpha$ anyon type is the renormalization of the quantum trace. This renormalization gives rise to a notion of the \textit{modified quantum dimensions}. For the $\alpha$ anyon type appearing in our framework, the usual quantum dimensions and the modified quantum dimensions, respectively, are  
\begin{align}
d_{\alpha} &= 0, & \mathbf{d}_{\alpha} &= -4\cfrac{q^\alpha - q^{-\alpha}}{q^{4\alpha} - q^{-4\alpha}} = -4\frac{\sin(\pi \alpha / 4)}{\sin ( \pi \alpha)}.
\end{align}

\subsection{Fusion}

The fusion rules for the $\alpha$ anyon type when the real parameter is $\alpha \in \R \setminus \Z$ are given in~\eqref{eq:alpha-fusion}. The fusion rules for when $\alpha \in \Z$ are more complicated, and we do not address them in this paper. The $\sigma$ and $\psi$ anyon types appearing in our framework admit the following fusion rules
\begin{equation} \label{eq:new-Ising-fusion}
\begin{split}
\psi \times \1 &= \psi,   \quad 
\sigma \times \1 = \sigma, \quad
\psi \times \psi = \1 + P_2, \\
\psi \times \sigma &= \sigma + S_{3/2}, \quad
\sigma \times \sigma = \1 + \psi.
\end{split}
\end{equation}

Notice that these fusion rules are an augmentation to the chiral Ising model fusion rules~\eqref{eq:Ising-fusion} where fusing $\psi\times \psi$ and $\psi \times \sigma$ now involve an additional anyon type, $P_2$ and $S_{3/2}$ respectively, of quantum trace zero. These anyons with quantum trace zero are removed during the semisimplification procedure used to get the standard theory of Ising anyons.

We define the \textit{fusion coefficients} $N^{c}_{ab}:=\dim (a,b)_c$ as the multiplicity of the anyon type $c$ appearing after fusing $a\times b$.  We say the labels $(a,b,c)$ are \textit{admissible} if $N^c_{ab}\neq 0$. These fusion coefficients $N^{c}_{ab}$  determine the admissibly of labeled trivalent fusion trees representing basis vectors in the local Hilbert spaces
\begin{equation} \label{eq:fusion-tree}
\begin{split}
\hackcenter{\begin{tikzpicture}[ scale=1.1]
  \draw[ultra thick, black] (0,0) to (.6,-.6) to (.6,-1.2);
  \draw[ultra thick, black] (1.2,0) to (.6,-.6);
   \node at (0,0.2) {$a$};
   \node at (1.2,0.2) {$b$};
   \node at (.8,-1) {$c$};
\end{tikzpicture} } \ .
\end{split}
\end{equation}
In our conventions, time travels upward.  
We assume all fusion trees are admissibly labeled and read from bottom to top.

\subsection{Inner Products} \label{sec:inner}

As a convention here, we take merge to be the dagger of split~(\ref{eq:fusion-tree}). 
\begin{equation} \label{eq:fusion-dagger}
\hackcenter{
\begin{tikzpicture}[ scale=1.1, yscale=-1.0]
  \draw[ultra thick, black] (0,0) to (.6,-.6) to (.6,-1.2);
  \draw[ultra thick, black] (1.2,0) to (.6,-.6);
   \node at (0,0.2) {$a$};
   \node at (1.2,0.2) {$b$};
   \node at (.6,-1.4) {$c$};
\end{tikzpicture} } :=
\left( \hackcenter{\begin{tikzpicture}[ scale=1.1]
  \draw[ultra thick, black] (0,0) to (.6,-.6) to (.6,-1.2);
  \draw[ultra thick, black] (1.2,0) to (.6,-.6);
   \node at (0,0.2) {$a$};
   \node at (1.2,0.2) {$b$};
   \node at (.8,-1) {$ c$};
\end{tikzpicture} }  \right)^{\dagger}  
\end{equation}
The definition of the dagger is explained in \cite{GLPMS}. 

The inner product of basis vectors~(\ref{eq-single-subit}) can be computed using the identity expressed in~\eqref{eq:bubble-pop} known as \textit{bubble pops}. 
\begin{equation}  \label{eq:bubble-pop}
\begin{split}
\hackcenter{\begin{tikzpicture}[ scale=1.1]
  \draw[ultra thick, black] (.6,-.6) to (.6,-1.2);
   \node at (.1,-.2) {$a$};
   \node at (1.1,-.2) {$b$};
   \node at (.8,-1) {$c$};
   \node at (.8,.6) {$c'$};
   \draw[ultra thick, black] (.9,-.2) to [out=270, in=45] (.6,-.6);
   \draw[ultra thick, black] (.9,-.2) to [out=90, in=315] (.6,.2);
   \draw[ultra thick, black] (.3,-.2) to [out=270, in=135] (.6,-.6);
   \draw[ultra thick, black] (.3,-.2) to [out=90, in=225] (.6,.2);
   \draw[ultra thick, black] (.6,.2) to (.6,.8);
\end{tikzpicture} }~=~\delta_{cc'}\text{sign}(B_{c}^{ab})~
\hackcenter{\begin{tikzpicture}[scale=1.1]
  \draw[ultra thick, black] (.6,-1.2) to (.6,.8);
   \node at (.8,-.2) {$c$};
\end{tikzpicture} } \ 
\end{split}
\end{equation}
The following values are the bubble pop data that appear in our encoding.
\begin{equation*} \label{eq:B-data}
\begin{aligned}
B^{\alpha\1}_{\alpha}=B^{\alpha\sigma}_{\alpha+1}=B^{\alpha\psi}_{\alpha+2}=B^{\sigma\sigma}_{\psi}=B^{\psi\sigma}_{S_{3/2}}=B^{\sigma\psi}_{S_{3/2}}=1,\\
B^{\sigma\sigma}_{\1}=(B^{\psi\sigma}_{\sigma})^{-1}=B^{\sigma\psi}_{\sigma}=-\sqrt{2},   \quad B_{\alpha-1}^{\alpha\sigma} = \cfrac{\sqrt{2}}{-1+\cot{\frac{\pi \alpha}{4}}},\\
B_{\alpha}^{(\alpha+2)\psi} = 2\cot{\frac{\pi \alpha}{4}},   \quad 
B_\alpha^{\alpha\psi} = \cfrac{\sqrt{2}\cos{\frac{\pi \alpha}{2}}}{1-\sin{\frac{\pi \alpha}{2}}},\\
B_{\alpha+1}^{\alpha S_{3/2}}=\frac{\sqrt{2}}{1-\tan\frac{\pi \alpha}{4}},   \quad B_{\alpha-1}^{\alpha S_{3/2}}=\frac{2+2\tan\frac{\pi\alpha}{4}}{-1+\cot\frac{\pi\alpha}{4}}.
\end{aligned}
\end{equation*}
These bubble pops are used to calculate the inner product on the computational space. For example,
\begin{equation}  \label{eq:basis-pop}
\hackcenter{\begin{tikzpicture}[scale=1.1]
  \draw[ultra thick, black] (.8,-0.8) to (.4,-.4) to (0,0) to (.4, .4) to (.8, .8) to (1.6,0) to (.8,-.8);
  \draw[ultra thick, black] (.4, .4) to (.8,0) to (.4,-.4);
  \draw[ultra thick, black] (.8, .8) to (.8, 1.2);
  \draw[ultra thick, black] (.8,-0.8) to (.8,-1.2);
   \node at (-0.2,0) {$\scriptstyle \alpha$};
   \node at (1.0,0) {$\scriptstyle \sigma$};
   \node at (1.8,0) {$\scriptstyle \sigma$};
   \node at (.35, .75) {$\scriptstyle \alpha \pm 1$};
   \node at (.35,-.75) {$\scriptstyle \alpha \pm 1$};
   \node at (1.0, 1.2) {$\alpha$};
   \node at (1.0,-1.2) {$\alpha$};
\end{tikzpicture} }
~=~\text{sign}(\mathbf{B}_{\alpha\pm 1})~
\hackcenter{\begin{tikzpicture}[scale=1.1]
  \draw[ultra thick, black] (.6,-1.2) to (.6,1.2);
   \node at (.8,-.2) {$\alpha$};
\end{tikzpicture} }  
\end{equation}
where $B_{\alpha\pm1}$ are given by \eqref{eq:Balphapmone}.  
The inner product $\braket{\cdot,\cdot}$ on $\mathcal{H}_1$ is defined as
\begin{equation}\label{eq:B-coefficients}
\left\langle 
\hackcenter{\begin{tikzpicture}[scale=0.8]
  \draw[ultra thick, black] (.8,-0.8) to (.4,-.4) to (0,0) ;
 \draw[ultra thick, black] (1.6,0) to (.8,-.8);
  \draw[ultra thick, black] (.8,0) to (.4,-.4);
  \draw[ultra thick, black] (.8,-0.8) to (.8,-1.2);
   \node at (-0.2,0) {$\scriptstyle \alpha$};
   \node at (1.0,0) {$\scriptstyle \sigma$};
   \node at (1.8,0) {$\scriptstyle \sigma$};
   \node at (.25,-.75) {$\scriptstyle x_1$};
   \node at (1.0,-1.2) {$\scriptstyle \alpha$};
\end{tikzpicture} } ,
\hackcenter{\begin{tikzpicture}[scale=0.8]
  \draw[ultra thick, black] (.8,-0.8) to (.4,-.4) to (0,0) ;
 \draw[ultra thick, black] (1.6,0) to (.8,-.8);
  \draw[ultra thick, black] (.8,0) to (.4,-.4);
  \draw[ultra thick, black] (.8,-0.8) to (.8,-1.2);
   \node at (-0.2,0) {$\scriptstyle \alpha$};
   \node at (1.0,0) {$\scriptstyle \sigma$};
   \node at (1.8,0) {$\scriptstyle \sigma$};
   \node at (.25,-.75) {$\scriptstyle x_2$};
   \node at (1.0,-1.2) {$\scriptstyle \alpha$};
\end{tikzpicture} } \right\rangle 
~=~\delta_{x_1x_2}\text{sign}(\mathbf{B}_{x_1})\mathbf{d}_{\alpha} 
\end{equation}
where $x_1,x_2\in \{\alpha\pm 1\}$.

More generally, the inner product on $\mathcal{H}_n$ can be computed by composing basis vectors with their adjoints and reducing to a multiple of a straight line using \eqref{eq:bubble-pop} and multiplying the result by the modified dimension $\mathbf{d}_{\alpha}$ and an overall sign.  
We can thus deduce that a computational $n$-qubit state $\ket{\phi}$ with $n_0$ 0s (corresponding to $\alpha + 1$) and $n_1$ 1s (corresponding to $\alpha - 1$) will have an inner product given by 
\begin{equation}\label{eq:computational_inner_product}
 \begin{split}
 \braket{\phi, \phi} &= \\ 
 & 
 (-1)^{n+1} \left(\mathrm{sign}(\mathbf{B}_{\alpha+1})\right)^{n_0} 
 \left(\mathrm{sign}(\mathbf{B}_{\alpha-1})\right)^{n_1} \mathbf{d}_{\alpha}.
     \end{split}  
\end{equation}
Since $\mathbf{B}_{\alpha\pm 1}$ are negative for $\alpha \in (2,3)$, we see that the computational space is always positive-definite. he factor of $(-1)^{n+1}$ comes from the fact that we multiply the Hermitian form by $(-1)$ for an even number of qubits.

\subsection{F-moves}

The \textit{6j-symbols} $(F^{abc}_d)_{nm}$, also commonly referred to as the \textit{$F$-symbols}, provide a change-of-basis between two fusion trees as shown in~\eqref{eq:F-move}.
\begin{equation}  \label{eq:F-move}
\begin{split}
\hackcenter{\begin{tikzpicture}[ scale=1.1]
  \draw[ultra thick, black] (0,0) to (.6,-.6) to (.6,-1.2);
  \draw[ultra thick, black] (0.6,0) to (.3,-.3);
  \draw[ultra thick, black] (1.2,0) to (.6,-.6);
   \node at (0,0.2) {$a$};
   \node at (.6,0.2) {$b$}; 
   \node at (1.2,0.2) {$c$};
   \node at (.8,-1) {$ d$};
   \node at (.3,-.6) {$\scriptstyle m$};
\end{tikzpicture} }
~=~\sum\limits_{n}(F_{d}^{abc})_{nm}
\hackcenter{\begin{tikzpicture}[ scale=1.1]
  \draw[ultra thick, black] (0,0) to (.6,-.6) to (.6,-1.2);
  \draw[ultra thick, black] (0.6,0) to (.9,-.3);
  \draw[ultra thick, black] (1.2,0) to (.6,-.6);
   \node at (0,0.2) {$a$};
   \node at (.6,0.2) {$b$}; 
   \node at (1.2,0.2) {$c$};
   \node at (.8,-1) {$ d$};
   \node at (.9,-.6) {$\scriptstyle n$};
\end{tikzpicture} } \ .
\end{split}
\end{equation}
Compatibility of the $F$-symbols require them to satisfy the {\em pentagon equation}  
\begin{equation} \label{eq:pentagon}
    (F^{ncd}_e)_{\ell m}(F^{abm}_e)_{np}
    = \sum_t (F^{abc}_\ell)_{nt} (F^{atd}_e)_{\ell r} (F^{bcd}_r)_{tm}~.
\end{equation}

We compile the $F$-symbols relevant to braiding in our encoding.  Note that to verify \eqref{eq:pentagon} one needs $F$-symbols involving the anyons $S_{3/2}$ and $P_2$.   First, we list the two one-dimensional data
\begin{align*}
(F^{\alpha \sigma\sigma}_{\alpha+2})_{\psi(\alpha+1)} &= 
1,\\
(F^{\alpha \sigma\sigma}_{\alpha-2})_{\psi(\alpha-1)} &= \mathrm{sign}(\sin\frac{\pi \alpha}{2}).
\end{align*}
For the $2\times 2$ $F$-matrices, one can compute the corresponding \textit{normalized}  $F$-symbol using the following formula
\begin{equation}
    (F^{abc}_d)_{nm}
    = \frac{\sqrt{B_{d}^{an}}\sqrt{B_{n}^{bc}}}{\sqrt{B_d^{mc}}\sqrt{B_{m}^{ab}}}(\tilde{F}^{abc}_d)_{nm}
\end{equation}
where the scalar involving bubble pops $\frac{\sqrt{B_{d}^{an}}\sqrt{B_{n}^{bc}}}{\sqrt{B_d^{mc}}\sqrt{B_{m}^{ab}}}$ applied to each matrix entry appropriately normalizes the matrix $\tilde{F}_{d}^{abc}$. These matrices are not manifestly (pseudo) unitary unless they are normalized.
\begin{align*}
\tilde{F}^{\alpha \sigma\sigma}_\alpha &= \left(\begin{array}{cc}
\tilde{F}_{\1(\alpha+1)} & \tilde{F}_{\1(\alpha-1)}\\
\tilde{F}_{\psi(\alpha+1)} & \tilde{F}_{\psi(\alpha-1)}
\end{array}\right)\\
&= \frac{1}{\sqrt{2} (q^{2\alpha} - 1)}\left(\begin{array}{cc}
q(q^{2\alpha} + q^2) & -(q^{2\alpha} - 1)\\
q^{2\alpha} - q^2 & q(q^{2\alpha} - 1)
\end{array}\right),\\
\tilde{F}^{\alpha\psi\sigma}_{\alpha+1} &= \left(\begin{array}{cc}
\tilde{F}_{\sigma\alpha} & \tilde{F}_{\sigma(\alpha+2)} \\
\tilde{F}_{S_{3/2}\alpha} & \tilde{F}_{S_{3/2}(\alpha+2)}
\end{array}\right)\\
&= \frac{1}{q^{2\alpha }+q^2}\left(\begin{array}{cc}
 (q^2-1)(q^{2\alpha }+q^2) & \left(q^2+1\right) \left(q^{2 \alpha }+1\right) \\
 (q^2+1)(q^{2\alpha }+q^2) &q^{2\alpha }-q^2
\end{array}\right),\\
\tilde{F}^{\alpha\psi\sigma}_{\alpha-1} &= \left(\begin{array}{cc}
\tilde{F}_{\sigma\alpha} & \tilde{F}_{\sigma(\alpha-2)} \\
\tilde{F}_{S_{3/2}\alpha} & \tilde{F}_{S_{3/2}(\alpha-2)}
\end{array}\right)\\
&= \frac{1}{q^{2 \alpha }-q^2}
\left(\begin{array}{cc}
 \left(q^2+1\right) \left(q^{2 \alpha }+q^2\right) & -2(q^{2 \alpha }-q^2) \\
 q^{2 \alpha }+1 & q^2(q^{2 \alpha }-q^2)
\end{array}\right),\\
\tilde{F}^{\alpha\sigma\psi}_{\alpha+1} &= \left(\begin{array}{cc}
\tilde{F}_{\sigma(\alpha+1)} & \tilde{F}_{\sigma(\alpha-1)} \\
\tilde{F}_{S_{3/2}(\alpha+1)} & \tilde{F}_{S_{3/2}(\alpha-1)}
\end{array}\right)\\
&= \frac{1}{q^{2 \alpha }-1}
\left(\begin{array}{cc}
 q^2 \left(q^{2 \alpha }+1\right) & -q(q^{2 \alpha }-1) \\
 \sqrt{2}\left(q^{2 \alpha }-q^2\right) & q^2(q^{2 \alpha }-1)
\end{array}\right),\\
\tilde{F}^{\alpha\sigma\psi}_{\alpha-1} &= \left(\begin{array}{cc}
\tilde{F}_{\sigma(\alpha+1)} & \tilde{F}_{\sigma(\alpha-1)} \\
\tilde{F}_{S_{3/2}(\alpha+1)} & \tilde{F}_{S_{3/2}(\alpha-1)}
\end{array}\right)\\
&= \frac{1}{q^{2 \alpha }-1}
\left(\begin{array}{cc}
 q \left(q^2+1\right) \left(q^{2 \alpha }+q^2\right) & -(q^{2 \alpha }-1) \\
 q^{2 \alpha }+1 & q(q^{2 \alpha }-1)
\end{array}\right).
\end{align*}

\subsection{Braiding}
 
The effect of exchanging two anyon types is determined from the \textit{$R$-move} shown below. 
\begin{equation}  \label{eq:R-move}
\begin{split}
\hackcenter{\begin{tikzpicture}[ scale=1.1]
  \draw[ultra thick, black] (.6,-.6) to (.6,-1.2);
   \node at (-.2,0.6) {$b$};
   \node at (1.4,0.6) {$a$};
   \node at (.8,-1) {$c$};
   \draw[ultra thick, black] (.9,-.15) to [out=270, in=90] (.6,-.6);
   \draw[ultra thick, black] (.9,-.15) to [out=90, in=-20] (0,.6);
   \path [fill=white] (.2,.2) rectangle (1,.4);
   \draw[ultra thick, black] (.3,-.15) to [out=270, in=90] (.6,-.6);
   \draw[ultra thick, black] (.3,-.15) to [out=90, in=200] (1.2,.6);
\end{tikzpicture} }
~=~R^{ba}_c
\hackcenter{\begin{tikzpicture}[ scale=1.1]
  \draw[ultra thick, black] (0,0) to (.6,-.6) to (.6,-1.2);
  \draw[ultra thick, black] (1.2,0) to (.6,-.6);
   \node at (0,0.2) {$b$};
   \node at (1.2,0.2) {$a$};
   \node at (.8,-1) {$ c$};
\end{tikzpicture} }  
\end{split}
\end{equation}
The \textit{$R$-symbols} are given as follows
\begin{alignat*}{2} \label{eq:R-data-Sung}
&R^{\alpha\psi}_{\alpha+2} = q^{3+\alpha},   \;\; &&
R^{\psi\alpha}_{\alpha+2} = q^{3+\alpha} ,\\
&R^{\alpha\sigma}_{\alpha+1} = q^{(3+\alpha)/2},   \;\; &&
R^{\sigma\alpha}_{\alpha+1} = q^{(3+\alpha)/2} ,\\
&R^{\alpha\psi}_{\alpha} = s_\alpha q^{1-\alpha},   \;\; &&
R^{\psi\alpha}_{\alpha} =s_\alpha q^{3+\alpha} ,\\
&R^{\alpha\sigma}_{\alpha-1} = s_\alpha q^{-(1+3\alpha)/2},   \;\; &&
R^{\sigma\alpha}_{\alpha-1} = s_\alpha q^{(7+\alpha)/2} ,\\
&R^{\alpha\psi}_{\alpha-2} = t_\alpha q^{1-3\alpha},   \;\; &&
R^{\psi\alpha}_{\alpha-2} = t_\alpha q^{5+\alpha} ,\\
&R^{\psi\sigma}_{S_{3/2}} = q,   \;\; &&
R^{\sigma\psi}_{S_{3/2}} = q ,\\
&R^{\psi\sigma}_{\sigma} = q,   \;\; &&
R^{\sigma\psi}_{\sigma} = q^3 ,\\
&R^{\sigma\sigma}_{\psi} = q^{1/2},    \;\; &&
R^{\sigma\sigma}_{\1} = q^{5/2}.
\end{alignat*} 
where
\begin{equation}
\begin{split}
 s_\alpha &:=\begin{cases} 
      +1, & \alpha \in (0,1) \cup (5,8) \setminus \mathbb{Z} \mod 8\\
      -1, & \alpha \in (1,5)\setminus \mathbb{Z} \mod 8
   \end{cases}
\end{split}
\end{equation}
and 
\begin{equation}
\begin{split}
t_\alpha &:=\begin{cases} 
      +1, & \alpha \in (0,1) \cup (2,4) \setminus \mathbb{Z} \mod 4\\
      -1, & \alpha \in (1,2) \setminus \mathbb{Z} \mod 4.
   \end{cases}
\end{split}
\end{equation}


\noindent \textbf{Acknowledgements}
The authors are grateful to Bertrand Patureau-Mirand, Hubert Saleur, Zhenghan Wang, and Paolo Zanardi for helpful conversations related to this project.  A.D.L., F.I., and S.K. are partially supported by NSF grants DMS-1902092 and DMS-2200419, the Army Research Office W911NF-20-1-0075, and the Simons Foundation collaboration grant on New Structures in Low-Dimensional Topology. S.K. is supported by the NSF Graduate Research Fellowship DGE-1842487. J.S. is partially supported by the NSF grant DMS-2401375, a Simons Foundation Travel Support Grant and PSC CUNY Enhanced Award 66685-00 54.

\noindent \textbf{Author Contributions}  F.I., S.K., J.S., and A.D.L. performed the research and wrote the manuscript.

\noindent \textbf{Competing Interests} The authors declare no competing interests.

\noindent \textbf{Correspondence and requests for materials} should be addressed to Aaron D. Lauda.

\end{document}